\documentstyle{article}

\makeatletter
\@twosidetrue  %  Defines twoside option.

\@options

\newcounter{volume}

\def\daterec#1{\def\@daterec{#1}}
\def\cislo#1{\def\@cislo{#1}}
\def\year#1{\def\@year{#1}
  \setcounter{volume}{#1} \addtocounter{volume}{-1950}
  \def\@volume{\arabic{volume}} }

\def\authori#1{\def\@authori{#1}}
\def\authorii#1{\def\@authorii{#1}}
\def\authoriii#1{\def\@authoriii{#1}}
\def\addressi#1{\def\@addressi{#1}}
\def\addressii#1{\def\@addressii{#1}}
\def\addressiii#1{\def\@addressiii{#1}}
\def\pagesfromto#1{\def\@pagesfromto{#1}}
\def\headauthor#1{\def\@headauthor{#1}}
\def\headtitle#1{\def\@headtitle{#1}}
\def\specialhead#1{\def\@specialhead{#1}}
\def\evidence#1{\def\@evidence{#1}}

\def\@normalsize{\@setsize\normalsize{12pt}\xpt\@xpt
\abovedisplayskip 10pt plus2pt minus5pt%
\belowdisplayskip \abovedisplayskip
\abovedisplayshortskip  \z@ plus3pt%
\belowdisplayshortskip  6pt plus3pt minus3pt%
\let\@listi\@listI}   % Setting of \@listi added 9 Jun 87

\def\small{\@setsize\small{11pt}\ixpt\@ixpt
\abovedisplayskip 8.5pt plus 3pt minus 4pt%
\belowdisplayskip \abovedisplayskip
\abovedisplayshortskip \z@ plus2pt%
\belowdisplayshortskip 4pt plus2pt minus 2pt
\def\@listi{\leftmargin\leftmargini %% Added 22 Dec 87
\topsep 4pt plus 2pt minus 2pt\parsep 2pt plus 1pt minus 1pt
\itemsep \parsep}}

\def\footnotesize{\@setsize\footnotesize{9.5pt}\viiipt\@viiipt
\abovedisplayskip 6pt plus 2pt minus 4pt%
\belowdisplayskip \abovedisplayskip
\abovedisplayshortskip \z@ plus 1pt%
\belowdisplayshortskip 3pt plus 1pt minus 2pt
\def\@listi{\leftmargin\leftmargini %% Added 22 Dec 87
\topsep 3pt plus 1pt minus 1pt\parsep 2pt plus 1pt minus 1pt
\itemsep \parsep}}
\def\scriptsize{\@setsize\scriptsize{8pt}\viipt\@viipt}
\def\tiny{\@setsize\tiny{6pt}\vpt\@vpt}
\def\large{\@setsize\large{14pt}\xiipt\@xiipt}
\def\Large{\@setsize\Large{18pt}\xivpt\@xivpt}
\def\LARGE{\@setsize\LARGE{22pt}\xviipt\@xviipt}
\def\huge{\@setsize\huge{25pt}\xxpt\@xxpt}
\def\Huge{\@setsize\Huge{30pt}\xxvpt\@xxvpt}

\normalsize  % Choose the normalsize font.

\if@twoside                 % Values for two-sided printing:
\else                       % Values for one-sided printing:
\fi
\advance\textheight by \topskip
\skip\footins 9pt plus 4pt minus 2pt  % Space between last line of text and
\@beginparpenalty -\@lowpenalty    % Before a list or paragraph environment.
\@endparpenalty   -\@lowpenalty    % After a list or paragraph environment.
\@itempenalty     -\@lowpenalty    % Between list items.

\def\part{\par               % New paragraph
   \addvspace{4ex}           % Adds vertical space above title.
   \@afterindentfalse        % Suppresses indent in first paragraph.  Change
   \secdef\@part\@spart}     % to \@afterindenttrue to have indent.

\def\@part[#1]#2{\ifnum \c@secnumdepth >\m@ne    % IF secnumdepth > -1
        \refstepcounter{part}                    %  THEN step part counter
        \addcontentsline{toc}{part}{\thepart     %       add toc line
        \hspace{1em}#1}\else                     %  ELSE add unnumbered line
      \addcontentsline{toc}{part}{#1}\fi         % FI
   { \parindent 0pt \raggedright
    \ifnum \c@secnumdepth >\m@ne   % IF secnumdepth > -1
      \Large \bf Part \thepart     %   THEN Print 'Part' and
      \par \nobreak                %          number in \Large boldface.
    \fi                            % FI
    \huge \bf                      % Select \huge boldface.
    #2\markboth{}{}\par }          % Print title and set heading marks null.
    \nobreak                       % TeX penalty to prevent page break.
    \vskip 3ex                     % Space between title and text.
   \@afterheading                  % Routine called after part and
    }                              %     section heading.

\def\@spart#1{{\parindent 0pt \raggedright
    \huge \bf
    #1\par}                         % Title.
    \nobreak                        % TeX penalty to prevent page break.
    \vskip 3ex                      % Space between title and text.
    \@afterheading                  % Routine called after part and
  }                                 %     section heading.

\def\section{\@startsection {section}{1}{\z@}{3.5ex plus 0.5ex minus
    .2ex}{2.3ex plus .2ex}{\centering\bf}}
\def\subsection{\@startsection{subsection}{2}{\z@}{3ex plus 1ex minus
   .2ex}{1.5ex plus .2ex}{\small\bf}}
\def\subsubsection{\@startsection{subsubsection}{3}{\z@}{3ex plus
-1ex minus -.2ex}{1.5ex plus .2ex}{\normalsize}}
\def\paragraph{\@startsection
     {paragraph}{4}{\z@}{-0em plus 0ex minus .2ex}{-2em}{\em}}
\def\subparagraph{\@startsection
     {subparagraph}{4}{\z@}{-0em plus 0ex minus
     .2ex}{-2em}{\normalsize}}

\def\appendix{\par
  \setcounter{section}{0}
  \setcounter{subsection}{0}
  \def\thesection{\Alph{section}}}

\def\thepart{\Roman{part}} % Roman numeral part numbers.
\def\thesection      {\arabic{section}}

\leftmargin\leftmargini
\labelwidth\leftmargini\advance\labelwidth-\labelsep

\def\@listI{\leftmargin\leftmargini \parsep 4pt plus 2pt minus 1pt%
\topsep 8pt plus 2pt minus 4pt%
\itemsep 4pt plus 2pt minus 1pt}

\let\@listi\@listI
\@listi

\def\@listii{\leftmargin\leftmarginii
   \labelwidth\leftmarginii\advance\labelwidth-\labelsep
   \topsep 4pt plus 2pt minus 1pt
   \parsep 2pt plus 1pt minus 1pt
   \itemsep \parsep}

\def\@listiii{\leftmargin\leftmarginiii
    \labelwidth\leftmarginiii\advance\labelwidth-\labelsep
    \topsep 2pt plus 1pt minus 1pt
    \parsep \z@ \partopsep 1pt plus 0pt minus 1pt
    \itemsep \topsep}

\def\@listiv{\leftmargin\leftmarginiv
     \labelwidth\leftmarginiv\advance\labelwidth-\labelsep}

\def\@listv{\leftmargin\leftmarginv
     \labelwidth\leftmarginv\advance\labelwidth-\labelsep}

\def\@listvi{\leftmargin\leftmarginvi
     \labelwidth\leftmarginvi\advance\labelwidth-\labelsep}

\def\theenumi{\arabic{enumi}}

\def\theenumii{\alph{enumii}}
\def\p@enumii{\theenumi}

\def\theenumiii{\roman{enumiii}}
\def\p@enumiii{\theenumi(\theenumii)}

\def\p@enumiv{\p@enumiii\theenumiii}

\def\verse{\let\\=\@centercr
  \list{}{\itemsep\z@ \itemindent -1.5em\listparindent \itemindent
          \rightmargin\leftmargin\advance\leftmargin 1.5em}\item[]}

\def\descriptionlabel#1{\hspace\labelsep #1}   %\bf odstraneno
\def\description{\list{}{\labelwidth\z@ \itemindent-\leftmargin
       \let\makelabel\descriptionlabel}}

\def\titlepage{\@restonecolfalse\if@twocolumn\@restonecoltrue\onecolumn
     \else \newpage \fi \thispagestyle{empty}\c@page\z@}

\def\endtitlepage{\if@restonecol\twocolumn \else \newpage \fi}

\tabbingsep \labelsep   % Space used by the \' command.  (See LaTeX manual.)

\skip\@mpfootins = \skip\footins

\def\@pnumwidth{1.55em}
\def\@tocrmarg {2.55em}
\def\@dotsep{4.5}
\def\tableofcontents{\section*{Contents\@mkboth{CONTENTS}{CONTENTS}}
  \@starttoc{toc}}

\def\l@part#1#2{\addpenalty{\@secpenalty}
   \addvspace{2.25em plus 1pt}  % space above part line
   \begingroup
   \@tempdima 3em         % width of box holding part number, used by
     \parindent \z@ \rightskip \@pnumwidth             %% \numberline
     \parfillskip -\@pnumwidth
     {\large \bf          % set line in \large boldface
     \leavevmode          % TeX command to enter horizontal mode.
     #1\hfil \hbox to\@pnumwidth{\hss #2}}\par
     \nobreak             % Never break after part entry
   \endgroup}

\def\l@section#1#2{\addpenalty{\@secpenalty}  % good place for page break
   \addvspace{1.0em plus 1pt}  % space above toc entry
   \@tempdima 1.5em            % width of box holding section number
   \begingroup
     \parindent \z@ \rightskip \@pnumwidth
     \parfillskip -\@pnumwidth
     \bf                  % Boldface.
     \leavevmode          % TeX command to enter horizontal mode.
      \advance\leftskip\@tempdima  %% added 5 Feb 88 to conform to
      \hskip -\leftskip            %% 25 Jan 88 change to \numberline
     #1\nobreak\hfil \nobreak\hbox to\@pnumwidth{\hss #2}\par
   \endgroup}

\def\l@subsection{\@dottedtocline{2}{1.5em}{2.3em}}
\def\l@subsubsection{\@dottedtocline{3}{3.8em}{3.2em}}
\def\l@paragraph{\@dottedtocline{4}{7.0em}{4.1em}}
\def\l@subparagraph{\@dottedtocline{5}{10em}{5em}}

\def\thebibliography#1{\section*{\small \bf References}
\small\list
  {[\arabic{enumi}]}{\settowidth\labelwidth{[#1]}\leftmargin\labelwidth
    \advance\leftmargin\labelsep
    \usecounter{enumi}}
    \def\newblock{\hskip .11em plus .33em minus .07em}
    \sloppy\clubpenalty4000\widowpenalty4000
    \sfcode`\.=1000\relax}

\newif\if@restonecol

\def\theindex{\@restonecoltrue\if@twocolumn\@restonecolfalse\fi
\columnseprule \z@
\columnsep 35pt\twocolumn[\section*{Index}]
    \@mkboth{INDEX}{INDEX}\thispagestyle{plain}\parindent\z@
    \parskip\z@ plus .3pt\relax\let\item\@idxitem}

\def\@idxitem{\par\hangindent 40pt}

\def\endtheindex{\if@restonecol\onecolumn\else\clearpage\fi}

\def\footnoterule{\kern-3\p@
  \hrule width .4\columnwidth
  \kern 2.6\p@}                 % The \hrule has default height of .4pt .

\newcounter{redfootnote}

\long\def\@makefntext#1{\parindent 1em\noindent
            \hbox to 1.8em{\hss$^{\@thefnmark}$)}\hspace{2pt}#1}

\long\def\@makefnmark{\hspace{0.5pt}\hbox{$^{\@thefnmark}$\hspace{-1pt})}}

\def\maketitle{\newpage 
\vskip-20pt
 \begingroup
 \def\thefootnote{\fnsymbol{footnote}}
 \def\@makefnmark{\hbox{%to 0pt
{$^{\@thefnmark}$)\hss}}}
 \newpage
 \global\@topnum\z@
  \@maketitle
 \thispagestyle{prvni}
 \endgroup
 \setcounter{redfootnote}{\value{footnote}}
 \setcounter{footnote}{0}
 \let\maketitle\relax
 \let\@maketitle\relax
 \gdef\@author{}\gdef\@title{}}

\def\@maketitle{\null
\vskip-17pt
{\centering
   {\bf \@title \par} \vskip 1em
   {\lineskip 2em  {\sc\@authori\par}}
   \medskip
   {\small \sl \@addressi\par}
   {\setbox\@tempboxa\hbox{\@authorii}
       \ifdim\wd\@tempboxa >0pt
          \vskip 1em {\sc\@authorii\par}
           \medskip
           {\small\sl\@addressii\par}
           {\setbox\@tempboxa\hbox{\@authoriii}
              \ifdim\wd\@tempboxa >0pt
                 \vskip 1em {\sc\@authoriii\par}
                  \medskip
                  {\small\sl\@addressiii\par}
              \else
              \fi}
       \else
       \fi}}
 \vskip 1em
  { \setbox\@tempboxa\hbox{}
   \ifdim \wd\@tempboxa >0pt   % IF longer than one line:
{\raggedleft \small \par}  %   THEN write.
     \else                        %   ELSE  nic.
          \fi}
 \par
 \vskip 1em
   }
\def\abstract{\small}

\mark{{}{}}   % Initializes TeX's marks

\def\rovnice#1{\def\@eqnnum{(#1)}}

\def\ps@prvni{\let\@mkboth\markboth
\def\@oddfoot{\hbox{}{}\hfill\rm\thepage}
\def\@evenfoot{\hbox{}\rm\thepage\hfill{}}
\def\@evenhead{}
\def\@oddhead{\hfill {\small LPTENS--00/31}}  }

\def\ps@normalni{\let\@mkboth\markboth
\def\@oddfoot{\hbox{}{}\hfil\rm\thepage}
\def\@evenfoot{\hbox{}\rm\thepage\hfil{}}
\def\@evenhead{\hbox{}\hfil}
\def\@oddhead{\hbox{}\hfil}  }

\def\ps@zcelaprvni{\let\@mkboth\markboth
\def\@oddfoot{\hbox{}{} \hfil\rm\thepage}
\def\@evenfoot{}
\def\@evenhead{}
\def\@oddhead{}  }

\def\ps@zvlastni{\let\@mkboth\markboth
\def\@oddfoot{\hbox{}{}\hfil\rm\thepage}
\def\@evenfoot{\hbox{}\rm\thepage\hfil{}}
\def\@evenhead{}
\def\@oddhead{} }

\def\ps@specialni{\let\@mkboth\markboth
\def\@oddfoot{}
\def\@evenfoot{\hbox{}\rm\thepage\hfil {}}
\def\@evenhead{\hbox{}\hfil\footnotesize\sl\@specialhead\hfil}
\def\@oddhead{}  }

\ps@normalni                   % 'plain' page style
\onecolumn                  % Single-column.
\if@twoside\else\fi % Ragged bottom unless twoside option.

\makeatother

\def\bc{\begin{center}}
\def\ec{\end{center}}
\def\bfl{\begin{flushleft}}
\def\efl{\end{flushleft}}
\def\bfr{\begin{flushright}}
\def\efr{\end{flushright}}
\def\bi{\begin{itemize}}
\def\ei{\end{itemize}}
\def\ben{\begin{enumerate}}
\def\een{\end{enumerate}}
\def\bd{\begin{description}}
\def\ed{\end{description}}
\def\be{\begin{equation}}
\def\ee{\end{equation}}
\def\bea{\begin{eqnarray}}
\def\eea{\end{eqnarray}}
\def\beb{\begin{eqnarray*}}
\def\eeb{\end{eqnarray*}}
\def\ba{\begin{array}}
\def\ea{\end{array}}
\def\bfg{\begin{figure}}
\def\efg{\end{figure}}
\def\bt{\begin{table}}
\def\et{\end{table}}
\def\btu{\begin{tabular}}
\def\etu{\end{tabular}}
\def\bbib#1{\begin{thebibliography}{#1}}
\def\ebib{\end{thebibliography}}

\def\slacs#1{\setlength{\arraycolsep}{#1}}

\def\I{{\rm i}}
\def\E{{\rm e}}
\def\D{{\rm d}}

\newfont{\cmbxix}{cmbx9}
\newfont{\cmbxviii}{cmbx8}
\newfont{\cmbxvii}{cmbx7}
\newfont{\cmbxvi}{cmbx6}
\newfont{\cmbxv}{cmbx5}

\newfont{\cmmibix}{cmmib9}
\newfont{\cmmibviii}{cmmib8}
\newfont{\cmmibvii}{cmmib7}
\newfont{\cmmibvi}{cmmib6}
\newfont{\cmmibv}{cmmib5}

\newfont{\cmbsyx}{cmbsy10}
\newfont{\cmbsyix}{cmbsy9}
\newfont{\cmbsyviii}{cmbsy8}
\newfont{\cmbsyvii}{cmbsy7}
\newfont{\cmbsyvi}{cmbsy6}
\newfont{\cmbsyv}{cmbsy5}

\begin{document}

\title{SEQUENTIAL BETHE VECTORS AND THE QUANTUM ERNST
SYSTEM\,\footnote{Presented at the 9th Colloquium ``Quantum groups and
integrable systems'', Prague, 22--24 June 2000}}
\authori{M. Niedermaier}
\addressi{Department of Physics, University of Pittsburgh, Pittsburgh,
  PA 15260, U.S.A.}  
\authorii{H. Samtleben}     %otherwise {}
\addressii{LPT-ENS, 24 Rue Lhomond, 75231 Paris C{\'e}dex 05,
France\,\footnote{UMR 8548: Unit{\'e} Mixte de Recherche du CNRS et de
  l'ENS.}}
\authoriii{}
\headtitle{Sequential Bethe vectors \ldots}
\headauthor{M. Niedermaier and H. Samtleben}  
%use "et al." for more than 3 authors
\specialhead{M. Niedermaier and H. Samtleben: Sequential Bethe vectors \ldots}

%%%%%%%%%%%%%%%%%%%%%%%%%%%%%%%%%%%%%%%%%%%%%%%%%%%%%%%%
\maketitle

\begin{abstract} 
  We give a brief review on the use of Bethe ansatz techniques to 
  construct solutions of recursive functional equations which 
  emerged in a bootstrap approach to the quantum Ernst system. 
  The construction involves two particular limits of a rational Bethe 
  ansatz system with complex inhomogeneities.  First, we pinch two 
  insertions to the critical value. This links Bethe systems 
  with different number of insertions and leads to the concept of 
  sequential Bethe vectors. Second, we study the semiclassical limit 
  of the system in which the scale parameter of the insertions tends 
  to infinity. 
\end{abstract} 
 
\slacs{.6mm}

\section{Functional equations for matrix elements in the quantum Ernst 
  system}  
 
In \cite{NieSam00} we proposed a bootstrap approach to describe 
the quantum theory descending from the Ernst equation of general 
relativity \cite{Erns68}. In upshot, the quantum theory is described 
in terms of matrix elements e.g.\ of the metric operator between 
spectral-transformed multi-vielbein configurations. Functional 
equations for these matrix elements were derived from an underlying 
quadratic algebra similar to the way the form factor axioms 
\cite{Smir92} may be derived from an underlying algebra \cite{Nied95}. 
Eventually the mathematical problem consists in finding 
sequences of vector-valued functions 
\vspace{-5mm}

\begin{equation} 
f_A(\theta):= f_{a_N\ldots a_1}(\theta_N,\ldots,\theta_1)\;,
\;\;\; N \geq N_0\,, 
\label{i2} 
\end{equation} 
obeying the following system of functional equations:     
\begin{eqnarray}  
{\cal T}(\theta_0|\theta)_A^B \,f_B(\theta) &=& 
{\tau}(\theta_0|\theta) f_A(\theta)\;, 
\label{Ia}\\[1ex] 
f_A(\theta) &=&
L_k(\theta_{k+1,k})_A^B \,f_B 
(\sigma_k\theta)\;, 
\label{Ib}\\[1ex] 
{\rm Res}_{\,\theta_{k+1} = \theta_k + \I\hbar}\,f_A(\theta)  
&=& 
\tau(\theta_k|p_k\theta)\,C_{a_{k+1}a_k}  \, 
f_{p_k A}(p_k \theta)\;, 
\label{IIa}\\[1ex] 
{\rm Res}_{\,\theta_{k+1} = \theta_k-\I\hbar}\, 
C^{a_{k+1}a_k}f_A(\theta)   
&=&
\tau(\theta_k\!-\!\I\hbar|p_k\theta)\,f_{p_kA}(p_k \theta)\;. 
\label{IIb} 
\end{eqnarray} 
Let us briefly describe these equations and the objects 
featuring. Throughout, we use the rational $R$-matrix 
\be 
R(\theta)_{ab}^{cd} := \frac{r(\theta)}{\theta- \I\hbar}\, 
\left(\theta\,\delta_a^c \delta_b^d - 
\I\hbar\, \delta_a^d 
\delta_b^c\right) 
\;, 
\ee 
where $r(\theta)$ satisfies 
$r(\theta)r(\theta\!-\!\I\hbar)=1-\I\hbar/\theta$. Equation (\ref{Ia})  
results in the diagonalization of the operator ${\cal T}$, for 
a fixed number of arguments $N$. It is basically the familiar 
transfer matrix   
\begin{equation}\label{T} 
{\cal T}(\theta_0|\theta)_A^B = 
\Gamma_b^a\, 
R_{c_N a_N}^{b \;b_N}(\theta_{N,0})\; 
R_{c_{N-1}a_{N-1}}^{c_N\; b_{N-1}}(\theta_{N-1,0})\;\ldots\; 
R_{a\; a_1}^{c_2b_1}(\theta_{1,0})\;. 
\end{equation} 
The matrix $\Gamma$ here denotes a traceless $SL(2,C)$ matrix
which for simplicity we assume to be diagonalized: $\Gamma := 
\mbox{diag}\,(\I,-\I)$. The operator ${\cal T}$ descends from the 
central quantum current \cite{ResSem90} in the Yangian double 
at the critical value of the central extension; the 
latter appears as part of the underlying quantized algebra of 
conserved charges \cite{KorSam98}. 
 
Equation (\ref{Ib}) describes the behavior of the eigenvectors $f_A$ 
under permutation of the arguments; $L_k$ acts as
\begin{equation}
L_k(\theta_{k+1,k})_A^B =
\delta_{a_N}^{b_N}\ldots 
R_{a_{k+1}a_k}^{b_kb_{k+1}}(\theta_{k+1,k})\ldots 
\delta_{a_1}^{b_1} \;,
\end{equation}
and $(\sigma_k\theta)=
(\theta_N,\ldots,\theta_{k},\theta_{k+1},\ldots,\theta_1)$, where by
$\theta_{kl}$ we denote the difference $\theta_k\!-\!\theta_l$.
Solutions to (\ref{Ia}) constructed by the Bethe ansatz can naturally
be made compatible with (\ref{Ib}).  In this note we mainly focus on
the recursive equations (\ref{IIa}) and (\ref{IIb}).  There, $C_{ab}$
denotes the $sl$(2) invariant antisymmetric tensor and we have adopted
the following notation for contraction: $p_k \theta =
(\theta_N,\ldots,\theta_{k+2},\theta_{k-1},\ldots,\theta_1)$, $p_k A=
(a_N,\ldots,a_{k+2},a_{k-1},\ldots,a_1)$. These two equations link
solutions of different eigenvector problems (\ref{Ia}), with $N$ and
$N\!-\!2$ arguments, respectively, under the pinching
$\theta_{k+1}\rightarrow\theta_k\pm\I\hbar$. For the eigenvalues
$\tau(\theta_0)$ they imply the compatibility condition
\be 
\tau(\theta_0|\theta)\Big|_{\theta_{k+1} =\theta_k \pm \I\hbar} =  
\tau(\theta_0|p_k\theta)\;.  
\label{t2} 
\ee 
 
In the following, we subsequently construct solutions to the equations
(\ref{Ia})--(\ref{IIb}). Equation (\ref{Ia}) amounts to
diagonalization of the transfer matrix, which is a well studied
problem and can be solved by Bethe ansatz techniques
\cite{Fadd95,KoBoIz93}. Joint solutions of equations (\ref{Ia}),
(\ref{Ib}) can be obtained from them by a symmetrization procedure.
Equations (\ref{IIa}), (\ref{IIb}) lead to the concept of
sequential Bethe vectors, connecting Bethe roots with different number
of arguments. From a technical viewpoint the latter might also offer a
new recursive approach (cf.\ \cite{KirRes86}) to issues like
completeness of the Bethe vectors. Finally, we discuss the
semi-classical limit $\hbar\rightarrow0$ of the solutions.  For
details and further references we refer to \cite{NieSam00}.
 
\section{Bethe ansatz} 
 
In this section we describe the solutions of equation (\ref{Ia}). The 
spectrum of ${\cal T}$ is conveniently organized by 
the $SO(2)$ symmetry that leaves ${\cal T}$ invariant: 
\be 
{\cal T}(\theta_0|\theta)_A^B  
\left(\mbox{$\sum_k$} \Gamma_k\right)_B^C = 
\left(\mbox{$\sum_k$} \Gamma_k\right)_A^B  
{\cal T}(\theta_0|\theta)_B^C \;, 
\ee 
with 
\begin{displaymath} 
(\Gamma_k)_A^B =  
\delta_{a_N}^{b_N}\dots \Gamma_{a_k}^{b_k} \dots\delta_{a_1}^{b_1} \;. 
\end{displaymath} 
The eigenvalue problem (\ref{Ia}) hence decomposes into 
decoupled sectors  
\be\label{edec} 
{\cal T}(\theta_0|\theta)_A^B \,f_{e;B}(\theta)  
= 
{\tau}_e(\theta_0|\theta) f_{e;A}(\theta)\;,\qquad 
(\Gamma_k)_A^B\,f_{e;B} = \I e\,f_{e;A}\;, 
\ee 
where $e= N, N\!-\!2, \ldots , -N\!+\!2, -N$, denotes the $SO(2)$ 
charge. We denote by $\pm$ the $sl$(2) indices in the ``charged'' 
basis of eigenvectors. For small $N$ the eigenvalue problem 
(\ref{edec}) can be solved by brute force but for generic $N$ it is 
useful to employ the standard techniques of the Bethe ansatz (see 
e.g.\ \cite{Fadd95,KoBoIz93}) and to parameterize the solutions in 
terms of the roots of the Bethe equations. 
 
Transferred to the present context this construction may be outlined 
as follows: Denote by $\Omega_A := 
\delta_{a_N}^+\,\dots\,\delta_{a_1}^+$ the lowest weight vector of 
the $N$-fold tensor product of the fundamental representation of 
$sl$(2). Following the Bethe Ansatz procedure, candidate eigenstates 
are generated from $\Omega$ by the repeated action of  
\be 
B(t|\theta)_A^B~:=~\Gamma_-^c\, 
R_{c_N a_N}^{+ \;b_N}(\theta_{N,0})\; 
R_{c_{N-1}a_{N-1}}^{c_N\; b_{N-1}}(\theta_{N-1,0})\;\ldots\; 
R_{c\; a_1}^{c_2b_1}(\theta_{1,0}) 
\;. 
\ee 
The matrix operators $B(t|\theta)$ are commuting for different values 
of $t$ and each $B(t_\alpha|\theta)$ lowers the $SO(2)$ charge $e$ of 
a candidate eigenstate of ${\cal T}$ by two units.  The candidate 
eigenstates can be made proper eigenstates by turning the parameters 
$t_{\alpha}$ into judiciously chosen functions of the $\theta_j$. In 
upshot one obtains eigenvectors 
\be\label{evec} 
w_{e}(\theta)=\prod_{\alpha=1}^{\Lambda}  
B(t_\alpha|\theta)\, \Omega \;,\qquad 
\Lambda:={\textstyle \frac 12}(N\!-\!e)\;, 
\ee 
with eigenvalues 
\be\label{tau} 
\tau_e(\theta_0|\theta) = 
\I\,\prod_\alpha  
{ \frac{\theta_0\!-\!t_\alpha\!+\!\I\hbar/2} 
{\theta_0\!-\!t_\alpha\!+\!3 \I\hbar/2}} 
\;\prod_j r^{-1}(\theta_{0j}) 
- 
\I\,\prod_\alpha  
{ \frac{\theta_0\!-\!t_\alpha\!+\!5\I\hbar/2} 
{\theta_0\!-\!t_\alpha\!+\!3 \I\hbar/2}} 
\;\prod_j r(\theta_{0j}\!+\!\I\hbar) \;, 
\label{ev} 
\ee 
where the Bethe roots $t_\alpha$ are solutions of the following  
Bethe Ansatz equations (BAE)  
\be\label{BAE} 
\prod_{j=1}^N \frac{\theta_j\!-\!t_\alpha\!-\!i\hbar/2} 
                   {\theta_j\!-\!t_\alpha\!+\!i\hbar/2}  = - 
\prod_{\beta\not=\alpha} \frac{t_\beta\!-\!t_\alpha\!-\!i\hbar} 
                              {t_\beta\!-\!t_\alpha\!+\!i\hbar}\;, 
 \qquad\alpha = 1,\ldots,\Lambda\;. 
\ee 
The only modification of the BAE as compared to the standard case 
$\Gamma=I$ is the sign on the r.h.s.\ which comes from the ratio of 
the eigenvalues of $\Gamma$. This seemingly innocent modification 
turns out to have nontrivial consequences in the classical limit 
$\hbar\rightarrow0$ to be described in section \ref{SCL}. 
 
Let us now return to the eigenvectors (\ref{evec}). Clearly any 
eigenvector is only determined up to multiplication by an arbitrary 
scalar function.  The Bethe eigenvectors as constructed by 
(\ref{evec}) will in general not obey the exchange relations 
(\ref{Ib}). However, it is not difficult to modify them so that they 
do. Due to the symmetry ${\tau}_e(\theta_0|\sigma\theta) = {\tau}_e 
(\theta_0|\theta),$ for all permutations $\sigma \in \Sigma_N$, a 
joint solution of (\ref{Ia}), (\ref{Ib}) can be obtained simply by 
symmetrizing with the $R$-matrix. In brief, for any given Bethe 
eigenvector (\ref{evec}) the product 
\be\label{product} 
f_{e;A}(\theta) \propto   
\prod_{k > l}\frac{\I\,\psi(\theta_{kl})}{\theta_{kl}^2 -
(\I\hbar)^2}\,  
w_{e;A}(\theta)\;,  
\ee 
solves both (\ref{Ia}) and (\ref{Ib}). Here, the function $\psi$ 
satisfies $\psi(\theta)= r(\theta) \psi(-\theta)$, and 
$\psi(-\theta)\psi(\theta\!-\!\I\hbar) = -1$ 
and is explicitly given by 
\begin{displaymath} 
\psi(\theta) = \tanh\frac{\pi \theta}{2\hbar } \,\exp\left\{ 
\I\,\int_0^{\infty} \frac{\D t}{t}  
\frac{\E^{-t/2} + \E^{-t}}{1 + \E^{-t}} \, 
\frac{\sin \frac{t}{2\hbar}(\I\hbar + 2\theta)}{\cosh \frac{t}{2}}  
\right\} 
\;. 
\end{displaymath} 
The proportionality sign in (\ref{product}) indicates that 
this eigenvector may still be multiplied with a scalar function 
$\phi_e(\theta)$ completely symmetric in $\theta_N, \ldots,\theta_1$. 
This freedom will mostly be fixed by further imposing the pinching 
equations (\ref{IIa}), (\ref{IIb}).

\section{Sequential Bethe roots and vectors}\label{SBV} 
Next let us examine the behavior of the Bethe ansatz equations and 
their solutions under pinching $\theta_{k+1} \rightarrow \theta_k \pm 
\I\hbar$ of the arguments. The relations 
(\ref{IIa}), (\ref{IIb}) imply that the $SO(2)$ charge $e$ of the 
eigenvectors is conserved under $\theta_{k+1} \rightarrow \theta_k \pm 
\I\hbar$, i.e. 
\be 
N\rightarrow N\!-\!2\;,\quad e\rightarrow e \;, 
\quad \Lambda\rightarrow\Lambda\!-\!1\;. 
\ee 
This suggests that the Bethe roots describing these (special) 
sequences of eigenvectors should likewise be related. Indeed, the BAE 
(\ref{BAE}) are consistent with the following $N\rightarrow N\!-\!2$ 
reduction of their solutions 
\be 
\label{tred} 
t_\Lambda(\theta)\Big|_{\theta_{k+1} = \theta_k \pm \I\hbar}  
= \theta_k 
\pm {\textstyle{\frac{1}{2}}}\I\hbar\;,\qquad 
t_\alpha(\theta)\Big|_{\theta_{k+1} = \theta_k \pm \I\hbar}  
= t_\alpha(p_k\theta)\;, 
\quad\mbox{for } \alpha<\Lambda\;. 
\ee 
Since the Bethe roots are symmetric in all $\theta_j$, it suffices to 
verify (\ref{tred}) for the $\theta_{k+1} = \theta_k + \I\hbar$ case. 
It is easy to check that with (\ref{tred}) the BAE (\ref{BAE}) for 
$\alpha<\Lambda$ reduce to the BAE with $N\!-\!2$ insertions for the 
$t_\alpha(p_k\theta)$. The equation for $\alpha = \Lambda$ is slightly 
more subtle as it requires to specify the limit in which the pinched 
configuration is approached.  Entering with the ansatz 
\bea 
\label{tlred} 
t_\Lambda(\theta) =  
\theta_k\!+\!\frac{\I\hbar}{2}+\delta /Z(\theta) + 
{\cal O}(\delta^2) \,, \qquad \mbox{for}\quad \theta_{k+1} =  
\theta_k\!+\!\I\hbar + \delta\;,  
\eea 
into the $\alpha = \Lambda$ BAE one obtains at order $\delta^0$ a
linear equation for $Z(\theta)$. This can be taken to define
$Z(\theta)$ and shows that the reduction rule for
$t_{\Lambda}(\theta)$ is consistent as $\delta \rightarrow 0$.
 
Of course, not every solution of the BAE will satisfy (\ref{tred}), in
fact the vast majority won't.  The argument shows however that under
the same genericity assumptions under which solutions exist at all,
there also exists at each recursion step $N -2 \mapsto N$ at least one
$\Lambda$-tuple of Bethe roots enjoying the property (\ref{tred}). We
call a solution of the BAE a ``sequential'' tuple of Bethe roots, if
all roots are distinct and satisfy (\ref{tred}).  To justify the
terminology one may easily verify that (\ref{tred}) with (\ref{tau})
implies the compatibility equation (\ref{t2}) for the eigenvalue
$\tau_e(\theta_0|\theta)$.
 
The corresponding eigenvectors will satisfy equations (\ref{IIa}),
(\ref{IIb}) up to a scalar function. The construction is completed by
determining the symmetric function $\phi_e(\theta)$ multiplying
(\ref{product}) such that (\ref{IIa}), (\ref{IIb}) are identically
satisfied; c.f.~\cite{NieSam00}.
 
\section{Semi-classical limit}\label{SCL} 
 
Finally we study the limit $\hbar\rightarrow0$ of the joint solutions 
of (\ref{Ia})--(\ref{IIb}).  In the context of the quantized 
Ernst system this corresponds to the semiclassical limit of the matrix 
elements. The limit of the transfer matrix ${\cal T}$ is given by: 
\bea 
{\cal T}(\theta_0|\theta)_A^B &=& 
\I\hbar\,\sum_k \frac{\Gamma_k}{\theta_{0k}} +  
(\I\hbar)^2\,\left( 
-\sum_k \frac{\Gamma_k}{2\theta^2_{0k}}+\sum_k  
\frac{H_k}{\theta_{0k}} 
\right) + {\cal O}(\hbar^3)\;, \label{semiT}\\[1ex] 
\mbox{with}\quad  
H_k &=& \sum_{l\not=k}\frac{\Omega_{kl}\, 
(\Gamma_k+\Gamma_l)}{\theta_{kl}} 
\;, \quad 
(\Omega_{kl})_A^B =  
\delta_{a_N}^{b_N}\dots  
(\delta_{a_k}^{b_l}\delta_{a_l}^{b_k}- 
{\textstyle\frac12}\delta_{a_k}^{b_k}\delta_{a_l}^{b_l}) 
\dots\delta_{a_1}^{b_1}\,.
\nonumber 
\eea 
This expansion is valid either as a formal power series in $\hbar$ or,
with a numerical $\hbar$, in the region ${\rm Im}\,\theta_{0k} \gg
\hbar$, ${\rm Im}\,\theta_{lk} \gg \hbar$, $l\neq k$, in order to
prevent a mixing of different powers of $\hbar$.  The absence of a
term of order $\hbar^0$ in (\ref{semiT}) is due to the tracelessness
of $\Gamma$ and distinguishes this case from the usual situation
$\Gamma=I$. The matrices $\Gamma_k$ and Hamiltonians $H_k$ form a
family of mutually commuting operators. Simultaneous diagonalization
of the $\Gamma_k$ yields eigenvectors with only one nonvanishing
component $(\epsilon_N,\ldots,\epsilon_1)$
\be 
w^{\rm cl}_{e;A}{}^{(\epsilon)} = 0 \quad 
\mbox{unless} \quad  (a_N,\ldots,a_1) =  
(\epsilon_N,\ldots,\epsilon_1)\;,\quad 
{\textstyle \sum_j}\epsilon_j=e \;. 
\label{w2cl} 
\ee 
On these eigenvectors the $H_k$ act diagonally. Thus the first terms 
in the semiclassical expansion of the eigenvalues $\tau$ are 
\bea 
\label{semit}  
\tau(\theta_0|\theta) &=& -\hbar\sum_k 
\frac{\epsilon_k}{\theta_{0k}} +\I \hbar^2 \Bigg(\frac{1}{2} \sum_k 
\frac{\epsilon_k}{\theta^2_{0k}} 
{}-\sum_{k\not=l\atop\epsilon_k=\epsilon_l} 
\frac{\epsilon_k}{\theta_{0k}\theta_{kl}}\Bigg) +{\cal O}(\hbar)^3\; 
\;.  
\eea 
This phenomenon can also be understood in terms of the Bethe ansatz. 
In the limit $\hbar \rightarrow 0$, the symmetry of the solutions of 
(\ref{BAE}) in $\theta_N,\ldots,\theta_1$ gets lost. Rather, the Bethe 
roots turn out to behave like    
\be 
t_{\alpha}(\theta) = \theta_{j(\alpha)} +  
(\I\hbar)^2 \, s_{\alpha}(\theta)  
+ {\cal O}(\hbar^3)\;, 
\label{rootcl} 
\ee  
for some $j(\alpha) \in \{1,\ldots,N\}$ with $j(\alpha) \neq j(\beta)$
for $\alpha \neq \beta$ and uniquely defined functions
$s_{\alpha}(\theta)$. Generally one can show that the Bethe roots
admit a power series expansion in $\hbar$ (in the region ${\rm
Im}\,\theta_{kl} \gg \hbar,\,k \neq l$) whose coefficients are
uniquely determined by the assignment $\alpha \rightarrow j(\alpha)$
in (\ref{rootcl}). The limiting behavior (\ref{rootcl}) drastically
differs from the standard case $\Gamma=I$, where no minus sign appears
in the r.h.s.\ of the BAE (\ref{BAE}) and the latter turn into an
identity for $\hbar \rightarrow 0$.  The eigenvector (\ref{w2cl})
corresponding to (\ref{rootcl}) is given by
\be  
w^{\rm cl}_{e;A}{}^{(\epsilon)}\;,\qquad\mbox{where}\quad 
\epsilon_k=\left\{ 
\begin{array}{rl} -&\mbox{if $k=j(\alpha)$ for some $\alpha$} \\ 
                  +&\mbox{otherwise} \end{array}\right. \;. 
\ee 
 
Summarizing, the semiclassical limit of the eigenvectors is given by  
\be\label{fsolcl} 
f_{e;A}(\theta) = \hbar^{\Lambda} \,f^{\rm cl}_{e;A}(\theta) 
+ {\cal O}(\hbar^{\Lambda+1})\;,\quad \mbox{with}\quad 
f^{\rm cl}_{e;A}(\theta)= 
\phi_e^{\rm cl}(\theta)\,w^{\rm cl}_{e;A}{}^{(\epsilon)} 
\;\prod_{k >l} \frac{1}{\theta_{kl}^2}\;. 
\ee 
As shown, this expansion refers to a fixed relative size of the 
variables $\theta_N,\ldots,\theta_1$, say $\theta_N > 
\ldots>\theta_1$. The results for other orderings then are compatible 
with the classical limit of the exchange relations in (\ref{Ib}), i.e. 
\be 
f^{\rm cl}_{e;A}(\theta) =  
f^{\rm cl}_{e;\sigma_k A} 
(\sigma_k\theta)\;. 
\label{Icl} 
\ee 
 
It remains the natural question for a classical counterpart of the 
recursive relations (\ref{IIa}), (\ref{IIb}), i.e.\ about 
commutativity of the two limits which we have described in this and 
the foregoing section. Indeed, it may be shown on the level of the 
Bethe roots that the classical limit of the $t_\alpha$ (\ref{rootcl}) 
and the pinching operation (\ref{tred}) commute in the relevant 
situations. The final relation is 
\be 
{\rm Res}_{\,\theta_{k+1} = \theta_k}\, 
f^{\rm cl}_{e;A}(\theta)  =
\,c_0\,C_{\epsilon_{k+1}\epsilon_k}\, 
f^{\rm cl}_{e;p_k A} 
(p_k \theta) 
\left(  
{\textstyle\sum_{j\not=k+1,k}} 
\frac{\epsilon_j}{\theta_{kj}}\right)\;, 
\label{clres2} 
\ee 
with a constant $c_0$. 
The last factor on the right hand side, when restricted to 
$\theta_{k+1} =\theta_k$ and $\epsilon_{k+1} =-\epsilon_k$, equals 
the leading term in the $\hbar$ expansion (\ref{semit}) of the 
transfer matrix eigenvalues. This is the consistency condition on 
(\ref{clres2}) analogous to (\ref{t2}). 
 
In summary, the solutions of the functional equations
(\ref{Ia})--(\ref{IIb}) admit a consistent semi-classical expansion.
The leading term (\ref{fsolcl}) of this expansion has, for a given
ordering of $\theta_{N},\ldots,\theta_1$, only one non-vanishing
component; different orderings being related by (\ref{Icl}). Further
these terms are themselves linked by the recurrence relation
(\ref{clres2}). It should be interesting to see whether these leading
terms have a direct interpretation in the classical theory.
 
\bigskip 
{\small The work of M.~N. was supported by NSF grant 
97-22097. The work of H.~S. was supported by EU contract 
ERBFMRX-CT96-0012.}

%\poslednisuda 
\end{document}